\documentclass[aps,prl,twocolumn,showpacs,preprintnumbers]{revtex4}%
\usepackage{amssymb}
\usepackage{graphicx}
\usepackage{dcolumn}
\usepackage{bm}
\usepackage{amsmath}
\usepackage{amsfonts}%
\begin{document}
\title{Cooperative Lifting of Spin Blockade in a Three-Terminal Triple Quantum Dot}
\author{Takashi Kobayashi}
\affiliation{NTT\,Basic\,Research\,Laboratories,\,NTT\,Corporation,\,3-1\,Morinosato-Wakamiya,\,Atsugi\,243-0198,\,Japan}
\author{Takeshi Ota}
\affiliation{NTT\,Basic\,Research\,Laboratories,\,NTT\,Corporation,\,3-1\,Morinosato-Wakamiya,\,Atsugi\,243-0198,\,Japan}
\author{Satoshi Sasaki}
\affiliation{NTT\,Basic\,Research\,Laboratories,\,NTT\,Corporation,\,3-1\,Morinosato-Wakamiya,\,Atsugi\,243-0198,\,Japan}
\author{Koji Muraki}
\affiliation{NTT\,Basic\,Research\,Laboratories,\,NTT\,Corporation,\,3-1\,Morinosato-Wakamiya,\,Atsugi\,243-0198,\,Japan}
\date{\today}

\begin{abstract}
We report measurements of multi-path transport through a triple quantum dot (TQD) in the few-electron regime using a GaAs three-terminal device with a separate lead attached to each dot. When two paths reside inside the transport window and are simultaneously spin-blockaded, the leak currents through both paths are significantly enhanced. We suggest that the transport processes in the two paths cooperate to lift the spin blockade. Fine structures in transport spectra indicate that different kinds of cooperative mechanisms are involved, depending on the details of the three-electron spin states governed by the size of exchange splitting relative to nuclear spin fluctuations. Our results indicate that a variety of correlation phenomena can be explored in three-terminal TQDs.

\end{abstract}

\maketitle

%\preprint{APS/123-QED}

%Force line breaks with \\

%\email{Second.Author@institution.edu}
%\altaffiliation[Also at ]{Physics Department, XYZ University.}%Lines break automatically or can be forced with \\

%It is always \today, today,
%but any date may be explicitly specified

%PACS, the Physics and Astronomy
%Classification Scheme.
%\keywords{Suggested keywords}%Use showkeys class option if keyword
%display desired
%\setlength{\baselineskip}{22.5pt}
%%%%%%%%%%%%%%%%%%%%%%%%%%%%%%%%%%%%%%%%%%%%
%% MAINMATTER
%%%%%%%%%%%%%%%%%%%%%%%%%%%%%%%%%%%%%%%%%%%%
%%%%%%%%%%%%%%%%%%%%%%%%%%%%%%%%%%%%%%%%%%%%%%%%%%%%%%%%%%%%%%%%%%%%%%%%%%%%%%
%%%%%%%%%%%%%%%%%%%%%%%%%%%%%%%%%%%%%%%%%%%%%%%%%
%%%%%%%%%%%%%%%%%%%%%%%%%%%%%%%%%%%%%%%%%%%%%%%%%%%%%%%%%%%%%%%%%%%%%%%%%%%%%%
%Introduction

%%%%%%%%%%%%%%%%%%%%%%%%%%%%%%%%%%%%%%%%%%%%%%%%%%%%%%%%%%%%%%%%%%%%%%%%%%%%%%

Electron spin in a semiconductor quantum dot (QD) is considered as a candidate for a solid-state qubit suitable for scalable quantum information processing \cite{Loss98PRA}. 
Previous studies on single and double QDs (DQDs) have demonstrated coherent manipulation of electron spins \cite{Petta05PRL, Koppens06Nature, Nowack07Science, Pioro-Ladriere08NPhys, Nadj-Perge10Nature, Nowack11Science, Brunner11PRL, Shulman12Science}. 
In these studies, the Pauli spin blockade (SB) \cite{Ono02Science}, which forbids two electrons with parallel spins to occupy the same QD, was used to readout the spin state via spin-charge conversion. 
The successful demonstration of spin control and readout in single and double QDs has raised interest in exploring larger systems involving more QDs. 
Triple quantum dot (TQD) systems, which represent the first test for scaling up, are attracting particular interest, triggered by the proposal of a novel spin qubit controllable by purely electrostatic means \cite{DiVincenzo00Nature, Medford13PRL, Medford13NNano}, and are being intensively studied experimentally \cite{Rogge08PhyE,Rogge08PRB,Rogge10PhyE,Seo13PRL,Gaudreau06PRL,Schroer07PRB,Granger10PRB,Gaudreau11NPhys,Amaha12PRB,Amaha13PRL,Busl13NNano,Braakman13APL,Braakman13NNano} and theoretically \cite{Saraga03PRL,Michaelis06EPL,Kuzmenko06PRL,Emary07PRB,Poltl09PRB,Trocha08JPCM,Delgado08PRL,Busl10PRB,Hsieh12PRB}.

%%%%%%%%%%%%%%%%%%%%%%%%%%%%%%%%%%%%%%%%%%%%%%%%%%%%%%%%%%%%%%%%%%%%%%%%%%%%%%

In a TQD, the charge degeneracy condition that allows for sequential tunneling through all three QDs is severely limited, which makes it difficult to explore the TQD's complex electronic structure via transport measurements \cite{Schroer07PRB,Granger10PRB,Busl13NNano,Braakman13APL,Braakman13NNano}. 
A three-terminal TQD, with a separate lead attached to each QD, not only eliminates this difficulty but also opens the way for studying the rich physics engendered by the presence of multiple current paths involving three QDs. 
Interesting phenomena such as separation of spin-entangled electron pairs \cite{Saraga03PRL} and various interference effects \cite{Michaelis06EPL,Kuzmenko06PRL,Emary07PRB,Poltl09PRB} are predicted. 
However, previous experiments on three-terminal TQDs have been performed only in the many-electron regime \cite{Rogge08PRB,Rogge08PhyE,Rogge10PhyE,Seo13PRL}. 
Thus, many of the electronic structures and associated spin dynamics remain unexplored, including coupling with nuclear spins (NSs), which could be investigated in the few-electron regime by using an SB \cite{Koppens05Science,Kobayashi11PRL}.

%%%%%%%%%%%%%%%%%%%%%%%%%%%%%%%%%%%%%%%%%%%%%%%%%%%%%%%%%%%%%%%%%%%%%%%%%%%%%%

In this Letter, we report transport measurements through a three-terminal GaAs lateral TQD in the few-electron regime. 
We study two-path transport, where electrons enter the center QD and leave from either the left or right one. 
Thus, the TQD system can be viewed as a splitter comprised of two sets of DQDs sharing the center QD. 
When only one of the two paths is allowed to enter the transport window, finite-bias transport through each path exhibits a bias triangle with an SB, characteristic of a DQD \cite{vanderWiel03RMP}. 
In contrast, when both paths are inside the transport window and simultaneously spin-blockaded, we find a correlative enhancement of transport for both DQDs. 
We explain this correlation by a mechanism in which the two DQDs cooperate to lift the SB. 
Detailed transport spectra show that different kinds of cooperative mechanisms operate, reflecting the size of the exchange splitting with respect to NS fluctuations, which modifies details of the spin configuration of the three-electron states. 
Our results indicate that a variety of correlation phenomena can be explored in three-terminal TQDs.

%%%%%%%%%%%%%%%%%%%%%%%%%%%%%%%%%%%%%%%%%%%%%%%%%%%%%%%%%%%%%%%%%%%%%%%%%%%%%%

\begin{figure}
[pb]
\begin{center}
\includegraphics{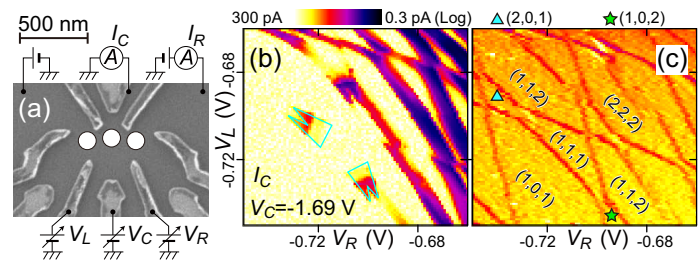}
\caption{(color online) (a) Scanning electron micrograph of the TQD sample. (b) $I_{C}$ as a function of $V_{R}$ and $V_{L}$ in the few-electron regime at $V_{C} = -1.69\ \text{V}$. (c) Charge detection signal measured simultaneously with (b).
}%
\end{center}
\end{figure}

%%%%%%%%%%%%%%%%%%%%%%%%%%%%%%%%%%%%%%%%%%%%%%%%%%%%%%%%%%%%%%%%%%%%%%%%%%%

The TQD used in this study was defined with Ti/Au gates in a GaAs/Al$_{0.3}$Ga$_{0.7}$As heterostructure containing a two-dimensional electron gas (density $2\times10^{15}$\thinspace m$^{-2}$) $95\ \text{nm}$ below the wafer surface [Fig.\thinspace1(a)]. 
The three QDs are serially arranged with finite tunnel coupling only between adjacent pairs, $t_{\ell c}$ ($t_{cr}$), for the left and center (center and right) QDs. 
Each QD has a separate electrical lead. 
The electrochemical potentials of the left and right leads are fixed at $-600\,\mu\text{eV}$, while that of the center lead is kept grounded. 
We measure the influx currents ($I_{R}$ and $I_{C}$) through the right and center leads as a function of gate biases $V_{L}$, $V_{C}$, and $V_{R}$. 
Current conservation allows us to deduce the influx current through the left lead as $I_{L}=-I_{R}-I_{C}$. 
The occupancies $(k,m,n)$ of the left, center, and right QDs are determined using a side-coupled quantum point contact charge sensor. 
Throughout this Letter, $k$ and $n$ are the actual electron numbers, while $m$ represents the effective number, which is smaller than the actual number by two. 
All measurements were performed with the sample mounted in a dilution refrigerator with a base temperature of $20$~$\,\text{mK}$.

%%%%%%%%%%%%%%%%%%%%%%%%%%%%%%%%%%%%%%%%%%%%%%%%%%%%%%%%%%%%%%%%%%%%%%%%%%%%%%

Figure\thinspace1(b) shows $I_{C}$ measured as a function of $V_{L}$ and $V_{R}$ at $V_{C}=-1.69\,\text{V}$.
The simultaneous charge detection reveals charge addition lines with three different slopes [Fig.\thinspace1(c)], indicating the formation of a TQD. 
DQD-like bias triangles \cite{vanderWiel03RMP} are clearly observed in $I_{C}$ around the charge degeneracy points between $(2,0,1)$ and $(1,1,1)$ and between $(1,1,1)$ and $(1,0,2)$. 
In the former, electrons pass through the DQD consisting of the center and left QDs (left DQD) with tunneling process $(1,1,1)\rightarrow (2,0,1)$. 
In the latter, electrons pass through the DQD consisting of the center and right QDs (right DQD) with tunneling process $(1,1,1)\rightarrow (1,0,2)$.

%%%%%%%%%%%%%%%%%%%%%%%%%%%%%%%%%%%%%%%%%%%%%%%%%%%%%%%%%%%%%%%%%%%%%%%%%%%

\begin{figure}
[ptb]
\begin{center}
\includegraphics{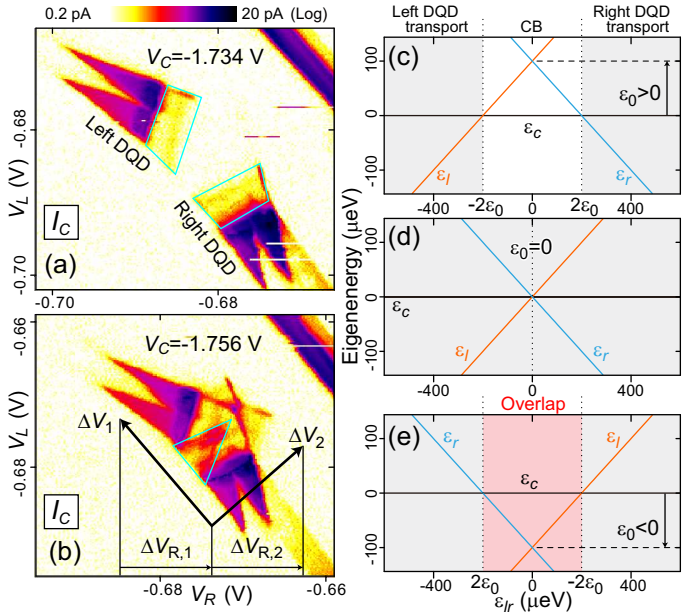}
\caption{(color online) (a), (b) $I_{C}$ vs $V_{R}$ and $V_{L}$ around the bias triangles of left and right DQDs, taken at $V_{C} = -1.734$ and $-1.756\ \text{V}$, respectively. (c)-(e) Energy level diagrams of $\varepsilon_{\ell}$, $\varepsilon_{c}$, and $\varepsilon_{r}$ as a function of $\varepsilon_{\ell r}$ for (c) $\varepsilon_{0}>0$, (d) $\varepsilon_{0}=0$, and (e) $\varepsilon_{0}<0$.
}%
\end{center}
\end{figure}

%%%%%%%%%%%%%%%%%%%%%%%%%%%%%%%%%%%%%%%%%%%%%%%%%%%%%%%%%%%%%%%%%%%%%%%%%%%

Figures\thinspace2(a) and (b) depict detailed measurements of $I_{C}$ near these bias triangles, taken at two slightly different $V_{C}$ values. 
For $V_{C}=-1.734\,\text{V}$ [Fig.\thinspace2(a)], the two bias triangles are separated. 
In the bottom part of each triangle, current is strongly suppressed by the DQD-like SB \cite{Koppens05Science}. 
A strikingly different behavior emerges when $V_{C}$ is slightly decreased ($V_{C}=-1.756\,$V) in such a way that the two bias triangles overlap each other [Fig.\thinspace2(b)]. 
Interestingly, $I_{C}$ in the overlapped region is larger than the simple sum of the currents in the non-overlapped regions.

%%%%%%%%%%%%%%%%%%%%%%%%%%%%%%%%%%%%%%%%%%%%%%%%%%%%%%%%%%%%%%%%%%%%%%%%%%%

The basic behavior of the bias triangles can be understood in terms of the relative alignment of the energy levels $(2,0,1)$, $(1,1,1)$, and $(1,0,2)$, which we denote by $\varepsilon_{\ell}$, $\varepsilon_{c}$ and $\varepsilon _{r}$, respectively. 
The important parameters are the detuning $\varepsilon _{\ell r}\equiv\varepsilon_{\ell}-\varepsilon_{r}$ between $\varepsilon_{\ell }$ and $\varepsilon_{r}$ and the relative alignment of their average with respect to $\varepsilon_{c}$, given by $\varepsilon_{0}\equiv\frac{1}{2}(\varepsilon_{\ell}+\varepsilon_{r})-\varepsilon_{c}$. 
Experimentally, $\varepsilon_{0}$ and $\varepsilon_{\ell r}$ are independently tunable by gate sweeps along the $\Delta V_{1}$ and $\Delta V_{2}$ axes defined in Fig.~2(b). 
The energy diagrams in Figs.\thinspace2(c)-(e) plot $\varepsilon_{\ell}$, $\varepsilon_{c}$, and $\varepsilon_{r}$ as a function of $\varepsilon_{\ell r}$ for (c) positive, (d) zero, and (e) negative $\varepsilon_{0}$. 
Transport through the left (right) DQD is governed by the $(2,0,1)$-$(1,1,1)$ [$(1,1,1)$-$(1,0,2)$] charge degeneracy point located at $\varepsilon_{\ell r}=-(+)2\varepsilon_{0}$, which we hereafter refer to as the left (right) DQD resonance. 
Energy conservation restricts transport through the left (right) DQD to the region $\varepsilon_{\ell r} \leq-2\varepsilon_{0}$ ($\varepsilon_{\ell r}\geq2\varepsilon_{0}$), where $\varepsilon_{\ell(r)}\leq\varepsilon_{c}$ is met. 
For $\varepsilon_{0}>0$, the transport regions of the two DQDs are separated by a region where both DQDs are in the Coulomb blockade [Fig.~2(c)]. 
This corresponds to the situation at $V_{C}=-1.734\,$V, where the bias triangles are separated [Fig.~2(a)]. 
In contrast, for $\varepsilon_{0}<0$, the transport regions of the two DQDs overlap [Fig.~2(e)], which leads to the overlapping bias triangles observed at $V_{C}=-1.756\,$V [Fig.~2(b)].

%%%%%%%%%%%%%%%%%%%%%%%%%%%%%%%%%%%%%%%%%%%%%%%%%%%%%%%%%%%%%%%%%%%%%%%%%%%%%%

\begin{figure}
[ptb]
\begin{center}
\includegraphics{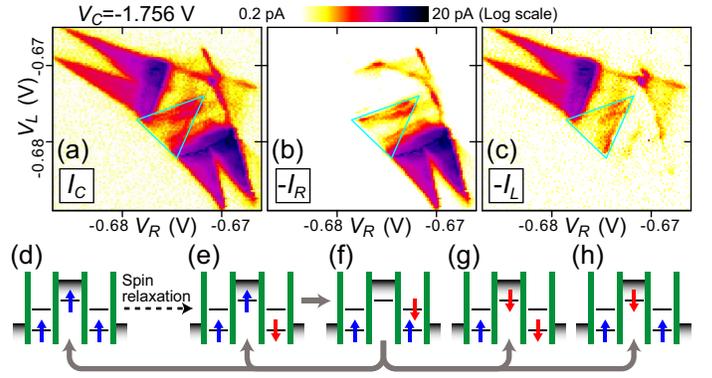}
\caption{(color online) (a) $I_{C}$, (b) $I_{R}$, and (c) $I_{L}$ vs $V_{R}$ and $V_{L}$ at $V_{C} = -1.756\ \text{V}$.  (d)-(h) Schematic illustration of processes involved in the cooperative lifting of the SB.
}%
\end{center}
\end{figure}

%%%%%%%%%%%%%%%%%%%%%%%%%%%%%%%%%%%%%%%%%%%%%%%%%%%%%%%%%%%%%%%%%%%%%%%%%%%%%%

The enhanced leak transport in the overlapped region is demonstrated more vividly by plotting $I_{R}$ and $I_{L}$ recorded simultaneously with $I_{C}$ [Figs.\thinspace3(a)-(c)]. 
Both $I_{R}$ and $I_{L}$ are clearly enhanced in the same gate-voltage region corresponding to the overlap. 
This clearly indicates that, although $I_{R}$ flows only through the right DQD, it is influenced by current flowing through the left DQD and vice versa.

%%%%%%%%%%%%%%%%%%%%%%%%%%%%%%%%%%%%%%%%%%%%%%%%%%%%%%%%%%%%%%%%%%%%%%%%%%%%%%

Before discussing the mechanism for the current enhancement, we introduce
notation for specifying TQD states. 
We use $\sigma_{\ell}$, $\sigma_{c}$, and $\sigma_{r}$ ($=\thinspace\uparrow,\downarrow$) to specify the spin of the singly occupied state in the left, center, and right QDs, respectively. 
For a doubly occupied state, we only consider a singlet $2S$, as the triplet lies much higher in energy. 
Thus, the states for the $(1,1,1)$ and $(2,0,1)$ charge configurations can be denoted by $|\sigma_{\ell},\sigma_{c},\sigma_{r}\rangle$ and $|2S,0,\sigma_{r}\rangle$, respectively. 
The SB in the non-overlapped region can be understood in the same way as that in a single DQD \cite{Ono02Science}. 
For example, transport via the $(1,1,1)\rightarrow (2,0,1)$ transition is blocked if the initial state $|\sigma_{\ell},\sigma _{c},\sigma_{r}\rangle$ has no overlap with $|(1,1)S,\sigma_{r}\rangle \equiv\frac{1}{\sqrt{2}}(\left\vert \uparrow,\downarrow,\sigma_{r} \right\rangle -\left\vert \downarrow,\uparrow,\sigma_{r}\right\rangle )$.

%%%%%%%%%%%%%%%%%%%%%%%%%%%%%%%%%%%%%%%%%%%%%%%%%%%%%%%%%%%%%%%%%%%%%%%%%%%%%%

We propose a mechanism for the observed current enhancement, which we refer to as cooperative lifting of the SB. 
As an example, we consider the situation where both DQDs are spin-blockaded by the occupation of $\left\vert \uparrow ,\uparrow,\uparrow\right\rangle $ [Fig.\thinspace3(d)]. 
Suppose that the SB is lifted by a spin flip in the right QD, resulting in $\left\vert \uparrow ,\uparrow,\downarrow\right\rangle $ [Fig.\thinspace3(e)]. 
Since $\left\vert \uparrow,\uparrow,\downarrow\right\rangle $ is orthogonal to $|(1,1)S,\sigma _{r}\rangle$ but not to $|\sigma_{\ell},(1,1)S\rangle$, a sequential tunneling through the right DQD $\left\vert \uparrow,\uparrow,\downarrow\right\rangle \rightarrow\left\vert \uparrow,0,2S\right\rangle \rightarrow\left\vert \uparrow,0,\sigma_{r}\right\rangle \rightarrow\left\vert \uparrow,\sigma _{c},\sigma_{r}\right\rangle $ follows. 
The spins of the electrons tunneling off from $\left\vert \uparrow,0,2S\right\rangle $ [Fig.\thinspace3(f)] or on to $\left\vert \uparrow,0,\sigma_{r}\right\rangle $ are random. 
Thus, the resultant state $\left\vert \uparrow,\sigma_{c},\sigma_{r}\right\rangle $ can take the four spin configurations shown in Figs.\thinspace3(d), (e), (g), and (h). 
It is seen that the SB is restored in (d), whereas the other three configurations can lead to successive sequential tunneling through the (e) right, (g) left, and (h) either DQD. 
It is noteworthy that the TQD state, which was orthogonal to $|(1,1)S,\sigma_{r}\rangle$ after the initial spin flip in the right QD [(e)], now has a finite overlap with $|(1,1)S,\sigma _{r}\rangle$ in (g) and (h) as a result of the SB's lifting and resultant transport in the right DQD. 
Such a sequence can occur independently of the initial SB state or in which QD the initial spin flip occurs. 
Thus, even if the SB in the right DQD is restored as in (g), the non-blockaded spin configuration in the left DQD leads to reloading into a $\left( 1,1,1\right)  $ state in which the SB in the right DQD is lifted. 
As a consequence of such cooperative effects, the lifting of the SB in either DQD induces a larger number of electrons to flow through both DQDs than in the case of independent DQDs.

%%%%%%%%%%%%%%%%%%%%%%%%%%%%%%%%%%%%%%%%%%%%%%%%%%%%%%%%%%%%%%%%%%%%%%%%%%%%%%

\begin{figure}
[ptb]
\begin{center}
\includegraphics{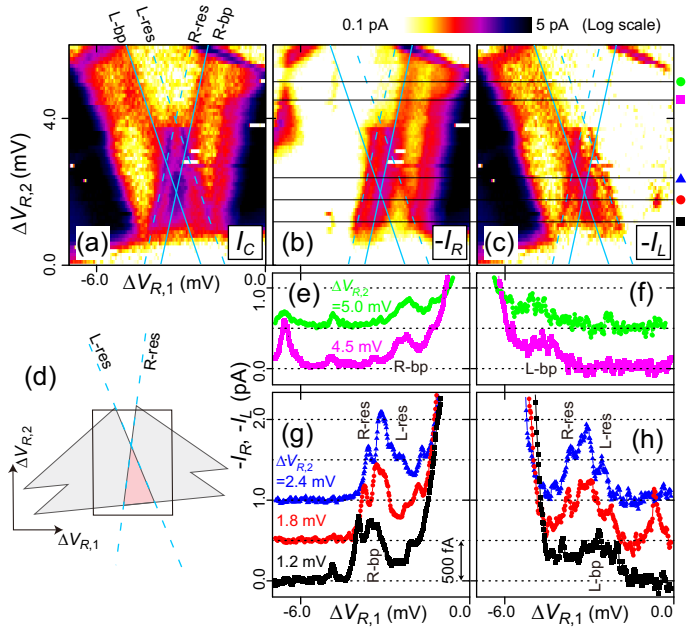}
\caption{(color online) (a)-(c) Detailed measurements of (a) $I_{C}$, (b) $I_{R}$, and (c) $I_{L}$ around the SB-SB overlap as a function of $\Delta V_{R,1}$ and $\Delta V_{R,2}$. (d) Schematic showing the correspondence between the region measured in (a)-(c) and the bias triangles. (e), (g) [(f), (h)] Slices of data in (b) [(c)] taken at $\Delta V_{R,2}$ values indicated by the horizontal lines in (b) and (c). (e) and (f) correspond to the non-overlapped region, and (g) and (h) to the overlapped region. The traces are offset vertically by $500\ \text{fA}$ for clarity.
}%
\end{center}
\end{figure}

%%%%%%%%%%%%%%%%%%%%%%%%%%%%%%%%%%%%%%%%%%%%%%%%%%%%%%%%%%%%%%%%%%%%%%%%%%%%%%

Deeper insights into the cooperative effects are provided by more detailed measurements around the SB-SB overlap [Fig.~4(a)-(c)]. 
As shown schematically in Fig.~4(d), the axes of Figs.~4(a)-(c) are taken nearly parallel to the $\Delta V_{1}$ and $\Delta V_{2}$ axes, with their scales projected onto the $V_{R}$ axis. 
The dashed lines labeled ``L-res'' and ``R-res'' mark the left- and right-DQD resonances, respectively. 
The data reveal that the enhanced leak transport is comprised of several fine structures. 
This is also seen in Figs.~4(e)-(h), where $I_{R}$ and $I_{L}$ along the five horizontal lines in Fig.~4(b) and (c) are displayed. 
First, we discuss the broad peaks running parallel to the right- and left-DQD resonance, indicated by the solid lines labeled ``L-bp'' and ``R-bp''. 
These peaks appear in both the overlapped and non-overlapped SB regions, but with different behavior. 
In the non-overlapped SB regions [Figs.\thinspace4(e) and (f)], the height of the peaks does not depend on$\text{ }\varepsilon_{0}$ (i.e., on $\Delta V_{R,2}$), as expected for independent DQD transport. 
The leak current of $\sim300\,$fA is consistent with that reported for a single DQD, which is known to arise from spin relaxation induced by an inhomogeneous NS field \cite{Koppens05Science,Jouravlev06PRL}. 
In contrast, inside the overlapped SB region [Figs.\thinspace 4(g) and (h)], the peak height varies with $\varepsilon_{0}$ by as much as a factor of $\sim2$, being enhanced up to $\sim1.0\,\text{pA}$.

%%%%%%%%%%%%%%%%%%%%%%%%%%%%%%%%%%%%%%%%%%%%%%%%%%%%%%%%%%%%%%%%%%%%%%%%%%%%%%

The evolution of the broad peak with $\varepsilon_{0}$ in the overlapped region is a manifestation of an interdependent relationship between the two DQDs in the cooperative transport. 
It is conceivable from the simplified picture in Figs.~3(d)-(h) that the transport through one DQD facilitates that through the other. 
Thus, one would expect the height of the peak in $I_{R}$ ($I_{L}$) to depend on the value of $I_{L}$ ($I_{R}$) at that gate voltage. 
Indeed, in Figs.~4(g) and (h), the broad peak in $I_{R}$ ($I_{L}$) grows as $\Delta V_{R,2}$ (and hence $\varepsilon_{0}$) increases and, accordingly, $I_{L}$ ($I_{R}$) at the corresponding gate voltage grows, reflecting the changes in the tuning parameters of the left (right) DQD.

%%%%%%%%%%%%%%%%%%%%%%%%%%%%%%%%%%%%%%%%%%%%%%%%%%%%%%%%%%%%%%%%%%%%%%%%%%%%%%

Another noticeable feature in Fig.~4 is the sharp peaks in $I_{R}$ and $I_{L}$
that appear along the R-res and L-res lines, respectively. 
These are distinct from the broad peaks discussed above in that transport occurs only with the help of the other DQD. 
Indeed, outside the overlap region, there is no transport feature visible along the R-res or L-res lines. 
In conventional DQDs, SB leak transport at resonance is suppressed by interdot tunnel coupling, which opens an anticrossing gap and pushes the singlet states out of the energy window accessible from the triplet via statistical fluctuations of the NS field $\Vert\Delta\vec{B}_{N}\Vert$~\cite{Koppens05Science}. 
In the present case, the absence of transport features along the R-res or L-res lines outside the overlap region indicates that both DQDs have large enough interdot tunnel coupling for such a situation to occur. 
Thus, the emergence of resonant leak transport in the overlap region suggests a non-trivial transport mechanism assisted by the other non-resonant DQD.

%%%%%%%%%%%%%%%%%%%%%%%%%%%%%%%%%%%%%%%%%%%%%%%%%%%%%%%%%%%%%%%%%%%%%%%%%%%%%%

\begin{figure}
[ptb]
\begin{center}
\includegraphics{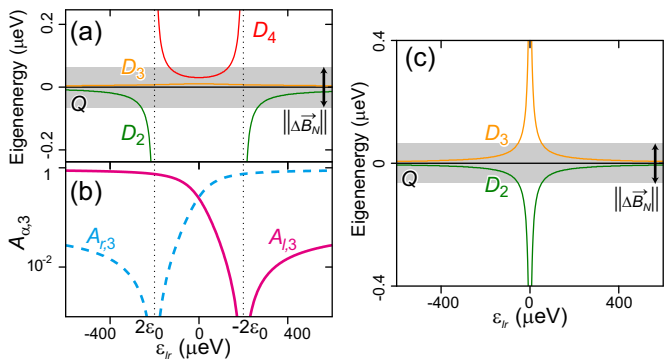}
\caption{(color online)
(a) Energy levels of $(1, 1, 1)$-like eigenstates calculated with $t_{\ell c} = t_{cr} = 1\,\mu\text{eV}$ for the same value of $\varepsilon_0$ ($< 0$) as in Fig.\thinspace2(e). (b) $A_{\ell,3}$ and $A_{r,3}$ for the same parameters as in (a). (c) Energies of eigenstates near the quadruplet ($Q$) calculated for $\varepsilon_{0}=0$.
}%
\end{center}
\end{figure} 

%%%%%%%%%%%%%%%%%%%%%%%%%%%%%%%%%%%%%%%%%%%%%%%%%%%%%%%%%%%%%%%%%%%%%%%%%%%%%%

To see this, we examine the nature of three-electron states in the presence of both $t_{\ell c}$ and $t_{cr}$. 
$|\sigma_{\ell},\sigma_{c},\sigma_{r}\rangle$, $|2S,0,\sigma_{r}\rangle$, and $|\sigma_{\ell},0,2S\rangle$ are hybridized to form four doublets (total spin $1/2$) and one quadruplet (total spin $3/2$), which we denote by $|D_{i}\rangle$ ($i=1,\cdots,4$) and $|Q\rangle$, respectively. 
Figure\thinspace5(a) shows their energy levels for $\varepsilon_{0}<0$, corresponding to the SB-SB overlap \cite{Supplement}. 
$|D_{2}\rangle$ and $|D_{4}\rangle$, which exhibit anticrossing, have finite $|2S,0,\sigma_{r}\rangle$ ($|\sigma_{l},0,2S\rangle$) components near the left-DQD (right-DQD) resonance. 
On the other hand, $|D_{3}\rangle$ as well as $|Q\rangle$ are constructed from $|\sigma_{\ell},\sigma_{c},\sigma_{r}\rangle$ only and thus have purely $(1,1,1)$-like charge configurations. 
The key to understanding the resonant lifting of SB is the unique property of $|D_{3}\rangle$, which is nearly degenerate with $|Q\rangle$. 
In this state, the two spins in the left DQD form a triplet near the left-DQD resonance, while those in the right DQD form a triplet near the right-DQD resonance. 
This is seen in the squared overlap integrals with the states comprising a $(1,1)$ singlet in the left or right DQD: $A_{\ell,3}\equiv|\langle(1,1)S,\sigma_{r}|D_{3} \rangle|^{2}$ and $A_{r,3}\equiv|\langle\sigma_{\ell},(1,1)S|D_{3}\rangle |^{2}$ [Fig.\thinspace5(b)]. 
$A_{\ell,3}$ ($A_{r,3}$) sharply drop near the left-DQD (right-DQD) resonances at $\varepsilon_{\ell r}=-2\varepsilon_{0}$ ($\varepsilon_{\ell r}=2\varepsilon_{0}$). 
The important observation is that $\,A_{r,3}$ remains finite when $A_{\ell,3}$ vanishes and vice versa. 
Therefore, although relaxation from $|Q\rangle$ to $|D_{3}\rangle$ does not directly contribute to SB leak current through the DQD on resonance, the occupation of $|D_{3}\rangle$ leads to sequential tunneling through the other (off-resonance) DQD, by which the system can be reloaded into $|D_{2}\rangle$ or $|D_{4}\rangle$. 
Note that these are resonant transport states of the on-resonance DQD, which accounts for the observed sharp peak along the resonance.

%%%%%%%%%%%%%%%%%%%%%%%%%%%%%%%%%%%%%%%%%%%%%%%%%%%%%%%%%%%%%%%%%%%%%%%%%%%%%%%

Finally, we note that current is suppressed at the crosspoint of the R-res and L-res lines [Figs.\thinspace4(a)-(c)]. 
Near such a double-resonance point, the effects of $t_{\ell c}$ and $t_{cr}$ are no longer separable. 
As shown in the energy diagram for $\varepsilon_{0}=0$ [Fig.\thinspace5(c)], near the resonance all the doublets $|D_{i}\rangle$ are split off from the quadruplets $|Q\rangle$, leaving no doublets available within a window of $\pm\Vert\Delta\vec {B}_{N}\Vert/2$. 
Thus, once the TQD is loaded into one of the $|Q\rangle$'s, the SBs in both DQDs are protected from NS fluctuations, and all three spins remain locked parallel to one another. 
Such an SB mechanism is distinct from that of the conventional SB in a DQD and is of genuine TQD nature.

%%%%%%%%%%%%%%%%%%%%%%%%%%%%%%%%%%%%%%%%%%%%%%%%%%%%%%%%%%%%%%%%%%%%%%%%%%%%%%%

In summary, we have demonstrated transport measurements through a three-terminal TQD in the few-electron regime. 
The SB-SB overlap brings out a correlation transport through the cooperation of two DQDs in the SB's lifting. 
Competition between exchange interaction and NS fluctuations leads to distinct cooperative mechanisms manifested by multiple peaks in the transport spectra. 
Our results show the potential of TQDs as a platform hosting a variety of correlation physics.

%%%%%%%%%%%%%%%%%%%%%%%%%%%%%%%%%%%%%%%%%%%%%%%%%%%%%%%%%%%%%%%%%%%%%%%%%%%%%%%

We thank Y. Tokura and T. Fujisawa for fruitful discussions and H. Murofushi for advice on device fabrication. 
This work was supported by Grant-in-Aid for Scientific Research (21000004).

%%%%%%%%%%%%%%%%%%%%%%%%%%%%%%%%%%%%%%%%%%%%%%%%%%%%%%%%%%%%%%%%%%%%%%%%%%%%%%%

\end{document}

\makeatletter 
\renewcommand{\figurename}{FIG.}
\renewcommand{\thefigure}{S\arabic{figure}}

%\nofiles

%\preprint{APS/123-QED}

\title{Supplementary Material for\\ ``Cooperative Lifting of Spin Blockade in a Three-Terminal Triple Quantum Dot''}% Force line breaks with \\
\author{Takashi Kobayashi}
\affiliation{NTT\,Basic\,Research\,Laboratories,\,NTT\,Corporation,\,3-1\,Morinosato-Wakamiya,\,Atsugi\,243-0198,\,Japan}
\author{Takeshi Ota}
\affiliation{NTT\,Basic\,Research\,Laboratories,\,NTT\,Corporation,\,3-1\,Morinosato-Wakamiya,\,Atsugi\,243-0198,\,Japan}
\author{Satoshi Sasaki}
\affiliation{NTT\,Basic\,Research\,Laboratories,\,NTT\,Corporation,\,3-1\,Morinosato-Wakamiya,\,Atsugi\,243-0198,\,Japan}
\author{Koji Muraki}
\affiliation{NTT\,Basic\,Research\,Laboratories,\,NTT\,Corporation,\,3-1\,Morinosato-Wakamiya,\,Atsugi\,243-0198,\,Japan}
\date{\today}
%\date{\today}% It is always \today, today,
             %  but any date may be explicitly specified

\maketitle
%\setlength{\baselineskip}{22.5pt}
%%%%%%%%%%%%%%%%%%%%%%%%%%%%%%%%%%%%%%%%%%%%
%% MAINMATTER
%%%%%%%%%%%%%%%%%%%%%%%%%%%%%%%%%%%%%%%%%%%%

%%%%%%%%%%%%%%%%%%%%%%%%%%%%%%%%%%%%%%%%%%%%%%%%%%%%%%%%%%%%%%%%%%%%%%%%%%%%%%%
\section{Hamiltonian}

%To calculate eigenstates $|D_{i}\rangle$ ($\,i=1,\cdots,4$) and $|Q\rangle$ without magnetic field, the $5 \times 5$ Hamiltonian $H_{\text{TQD}}^{S_z=+1/2}$ of the subspace where $z$-projection of the total spin is $+1/2$ is sufficient:
Eigenstates $|D_{i}\rangle$ ($\,i=1,\cdots,4$) and $|Q\rangle$ without magnetic field are calculated with $5 \times 5$ Hamiltonian $H_{\text{TQD}}^{S_z=+1/2}$ of the subspace where $z$-projection of the total spin is $+1/2$:
\begin{eqnarray*}
H_{\text{TQD}}^{S_z=+1/2} 
	&=&(\varepsilon_{\ell}-\varepsilon_{c})| 2S,0,\uparrow \rangle \langle 2S,0,\uparrow|
	+(\varepsilon_{r}-\varepsilon_{c})| \uparrow,0,2S \rangle \langle \uparrow,0,2S|\\
	& & -t_{\ell c}(\,| \downarrow,\uparrow,\uparrow \rangle\langle 2S,0,\uparrow |+| 2S,0,\uparrow \rangle\langle \downarrow,\uparrow,\uparrow |\,)
	+t_{\ell c}(\,| \uparrow,\downarrow,\uparrow \rangle\langle 2S,0,\uparrow |+| 2S,0,\uparrow \rangle\langle \uparrow,\downarrow,\uparrow |\,)\\
	&&+t_{cr}(\,| \uparrow,\downarrow,\uparrow \rangle\langle \uparrow,0,2S |+| \uparrow,0,2S \rangle\langle \uparrow,\downarrow,\uparrow |\,)
	-t_{cr}(\,| \uparrow,\uparrow,\downarrow \rangle\langle \uparrow,0,2S |+| \uparrow,0,2S \rangle\langle \uparrow,\uparrow,\downarrow |\,)\\
	&=&\left(
	\begin{array}{c|ccc|c}
	\varepsilon_{\ell r}/2+\varepsilon_{0} & -t_{\ell c} & t_{\ell c} & 0 & 0 \\
	\hline
	-t_{\ell c} & 0 & 0 & 0 & 0 \\
	t_{\ell c} & 0 & 0 & 0 & t_{cr} \\
	0 & 0 & 0 & 0 & -t_{cr} \\
	\hline
	0 & 0 & t_{cr} & -t_{cr} & -\varepsilon_{\ell r}/2+\varepsilon_{0} \\
	\end{array}
	\right) ,
\end{eqnarray*}
on a basis set of $| 2S,0,\uparrow \rangle$, $| \downarrow,\uparrow,\uparrow \rangle$, $| \uparrow,\downarrow,\uparrow \rangle$, $| \uparrow,\uparrow,\downarrow \rangle$, and $| \uparrow,0,2S \rangle$.

%%%%%%%%%%%%%%%%%%%%%%%%%%%%%%%%%%%%%%%%%%%%%%%%%%%%%%%%%%%%%%%%%%%%%%%%%%
%%%%%%%%%%%%%%%%%%%%%%%%%%%%%%%%%%%%%%%%%%%%%%%%%%%%%%%%%%%%%%%%%%%%%%%%%%

%%%%%%%%%%%%%%%%%%%%%%%%%%%%%%%%%%%%%%%%%%%%%%%%%%%%%%%%%%%%%%%%%%%%%%%%%%
%%%%%%%%%%%%%%%%%%%%%%%%%%%%%%%%%%%%%%%%%%%%%%%%%%%%%%%%%%%%%%%%%%%%%%%%%%